\documentclass[12pt]{iopart}
\usepackage{graphicx}
\usepackage[none]{harvard}

\begin{document}

\title[Transverse, Propagating Velocity Perturbations in Solar Coronal Loops]{Transverse, Propagating Velocity Perturbations in Solar Coronal Loops}

\author{I. De Moortel$^1$, D.J.~Pascoe$^2$, A.N.~Wright$^1$ and A.W.~Hood$^1$}

\address{$^1$ University of St Andrews, School of Mathematics \& Statistics, North Haugh, St Andrews, KY16 9SS}
\address{$^2$ Centre for Fusion, Space and Astrophysics, Department of Physics, University of Warwick, CV4 7AL}

\ead{ineke.demoortel@st-andrews.ac.uk}
\vspace{10pt}
\begin{indented}
\item[]June 2015 
\end{indented}

\begin{abstract}
As waves and oscillations carry both energy and information, they have enormous potential as a plasma heating mechanism and, through seismology, to provide estimates of local plasma properties which are hard to obtain from direct measurements. Being sufficiently near to allow high-resolution observations, the atmosphere of the Sun forms a natural plasma laboratory. Recent observations have revealed that an abundance of waves and oscillations is present in the solar atmosphere, leading to a renewed interest in wave heating mechanisms. 

This short review paper gives an overview of recently observed transverse, propagating velocity perturbations in coronal loops. These ubiquitous perturbations are observed to undergo strong damping as they propagate. Using 3D numerical simulations of footpoint-driven transverse waves propagating in a coronal plasma with a cylindrical density structure, in combination with analytical modelling, it is demonstrated that the observed velocity perturbations can be understood in terms of coupling of different wave modes in the inhomogeneous boundaries of the loops. Mode coupling in the inhomogeneous boundary layers of the loops leads to the coupling of the transversal (kink) mode to the azimuthal (Alfv\'en) mode, observed as the decay of the transverse kink oscillations. Both the numerical and analytical results show the spatial profile of the damped wave has a Gaussian shape to begin with, before switching to exponential decay at large heights. In addition, recent analysis of CoMP ({\it Coronal Multi-channel Polarimeter}) Doppler shift observations of large, off-limb, trans-equatorial loops shows that Fourier power at the apex appears to be higher in the high-frequency part of the spectrum than expected from theoretical models. This excess high-frequency FFT power could be tentative evidence for the onset of a cascade of the low-to-mid frequency waves into (Alfv\'enic) turbulence. 
\end{abstract}

%
%
%
%
%


\section{Introduction}

Waves and oscillations have long been considered as a possible mechanism to heat the solar corona but, with the advent of high resolution and high cadence observations, have also been considered as a viable tool to reveal information about local plasma properties through coronal (atmospheric) seismology. There is a large body of literature on both MHD wave heating and coronal seismology and we refer the interested reader to recent reviews on heating (e.g.~\cite{Klimchuk2006}, \cite{reale2010}, \cite{Parnell2012}) and seismology (e.g.~\cite{Nakariakov05}, \cite{Banerjee2007}, \cite{Taroyan2009}, \cite{IDM2012}, \cite{Arregui2015}).

Recent observations by \cite{Tomczyk2007} found ubiquitous perturbations travelling along coronal loops observed by the ground based coronagraph CoMP. The perturbations were observed as Doppler velocity perturbations travelling along the coronal loops, with speeds of the order of 600 km/s (\cite{Tomczyk2009}).  As no corresponding perturbations in the imaging observations were found, the authors interpreted the observed disturbances as incompressible. (Note however that the absence of intensity perturbations does not necessarily imply that the disturbances are incompressible - see e.g.~\cite{Cooper2003}, \cite{Gruszecki2012} or \cite{Antolin2013}.) Combining this incompressible nature with the very high propagation speed, lead \cite{Tomczyk2007} to interpret the observed perturbations as propagating Alfv\'en waves.  Similar transverse, propagating oscillations, interpreted as ``Alfv\'enic'' perturbations, have been observed along a variety of structures such as spicules (\cite{DePontieu2007}, \cite{He2009a}, \cite{He2009b}), X-ray jets (\cite{Cirtain2007}), prominence threads (\cite{Okamoto2007}) and coronal loops (\cite{Tomczyk2007}, \cite{Tomczyk2009}, \cite{McIntosh2011}).

The interpretation of these transverse perturbations as ``Alfv\'enic''  is not universally accepted. For example, \cite{TVD2008} argue that the waves observed by \cite{Tomczyk2007} should be interpreted as kink waves (due to the presence of flux tubes in the solar corona) rather than Alfv\'en waves (applicable to a uniform medium). If one interprets ``Alfv\'enic'' as waves dominated by magnetic tension as the restoring force, kink waves in a low-beta plasma and Alfv\'en waves in a homogeneous medium can both be considered to be special cases of Alfv\'enic waves. However, the interpretation in terms of Alfv\'en or kink waves could have considerable implications for the energy budget (see also \cite{Goossens2013} and \cite{TVD2014}) and possibly for the seismologically inferred magnetic field strength. Still, even with this uncertainty in mind, it is instructive to look at the energy budget estimated from the observations. Both \cite{DePontieu2007} and \cite{McIntosh2011} estimated that sufficient energy is available in the observed Alfv\'enic waves to heat the quiet-Sun corona. A much lower energy budget was reported by \cite{Tomczyk2007}, who report that the wave flux estimated in the observed off-limb coronal loops is several orders of magnitude too small to account for the heating of the local plasma. However, it is likely that the value reported by these latter authors is very much a lower limit, due to superposition effects along the line-of-sight and the low spatial resolution of CoMP (\cite{McIntosh2012}). Indeed, 3D numerical simulations by \cite{IDM-Pascoe2012} showed that the energy budgets estimated from observed, line-of-sight integrated Doppler velocities will be at least an order of magnitude smaller than the energy present in the 3D coronal volume.

Of particular interest to this review is the very rapid damping of the observed propagating, transverse waves, estimated  by \cite{Tomczyk2009}. These authors found the power contained in outward propagating waves to be substantially larger than the inward power. In other words, waves generated at one loop footpoint undergo substantial amplitude decay as they travel along the loops and are not observed to propagate down the opposite loop leg. In order to model the observed rapid damping and clarify the interpretation in terms of MHD modes, \cite{Pascoe2010} set up 3D numerical simulations of a generic footpoint displacement of a coronal flux tube. These authors found that the transverse density structuring (where there loop is assumed to be denser than the external plasma) leads to an intrinsic coupling between the bulk transverse displacement (a fast kink wave) and azimuthal ($m=1$) Alfv\'en waves. This ``mode coupling''  as the kink waves propagate along the loop is analogues to the process of resonant absorption in standing modes (see a review by e.g.~\cite{Goossens2011}). The process of mode coupling is a well studied topic in the context of the Earth's magnetosphere. Due to the restricted length of this review, we refer the interested reader to the review by  \cite{Wright2006}. 

It is important to note here that observed wave damping in coronal loops does not automatically imply dissipation, and hence heating. Indeed, although mode coupling may lead to observed wave damping due to the transfer of energy to the azimuthal Alfv\'en wave, it does not necessarily imply wave dissipation on the appropriate temporal and spatial scales. Of course, other effects such as viscous and resistive damping could also lead to (additional) wave damping and direct dissipation (see e.g.~the 2D linear, resistive slab model of \cite{Ofman1995}, who studied the heating and propagation of fast and shear Alfv\'en waves in coronal holes). Dissipation might be significantly enhanced due to the presence of background turbulence (e.g.~\cite{Hollweg1988} or \cite{Ofman2002}).

The results presented in this paper are a summary of the mode coupling studies described in detail in the series of papers \cite{Pascoe2010}, \cite{Pascoe2011}, \cite{Pascoe2012}, \cite{Pascoe2013}, \cite{Hood2013}, \cite{Pascoe2015}. In Section 2, the basic numerical setup of the 3D numerical mode coupling simulations is presented, followed by the results in Section 3 and a discussion of the Gaussian damping with height of the kink waves in Section 4. Section 5 contains a brief discussion and conclusions.

\section{Basic Numerical Setup}

\begin{figure}[t]
\centering{
\scalebox{.18}{\includegraphics{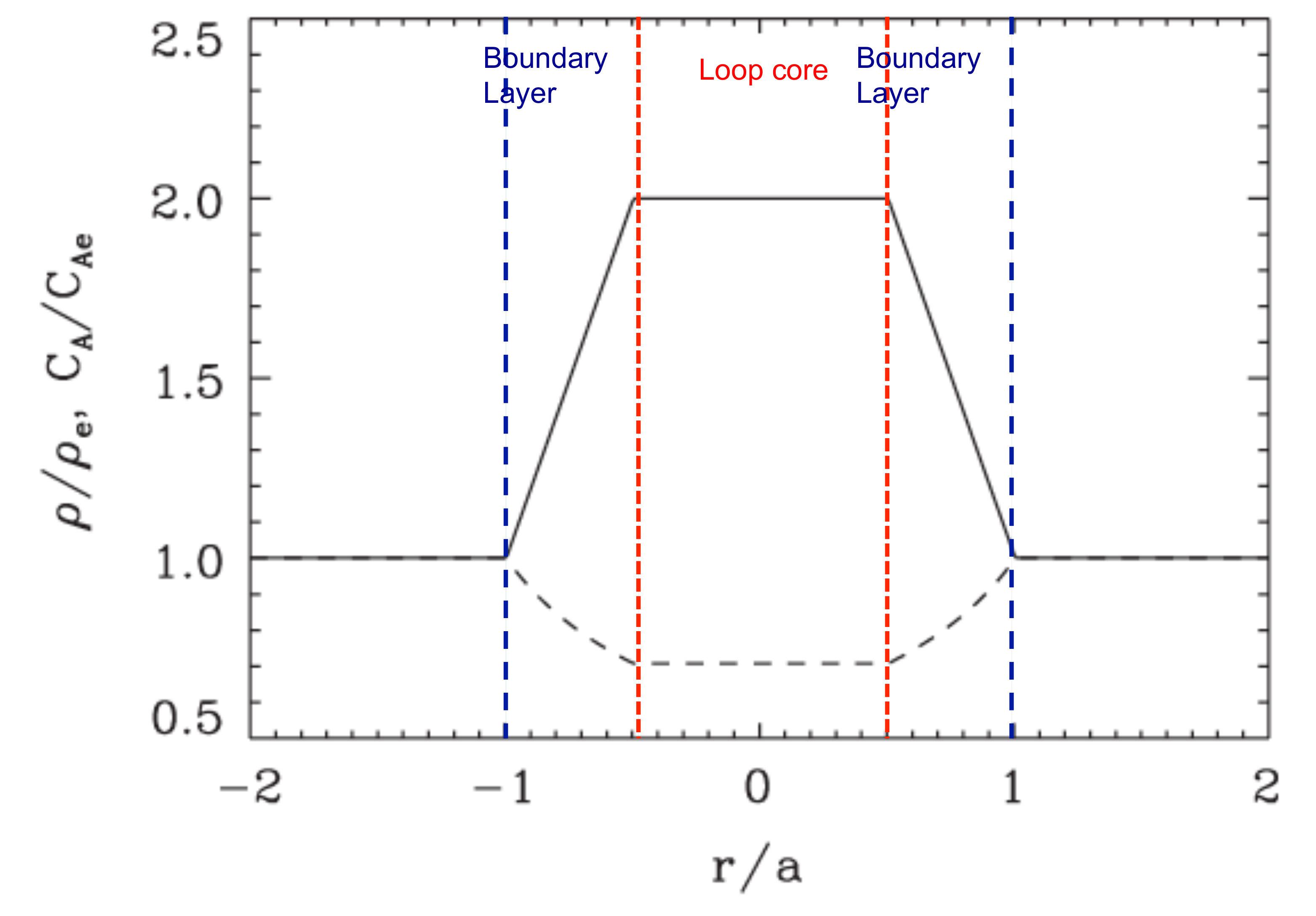}}
\scalebox{.5}{\includegraphics{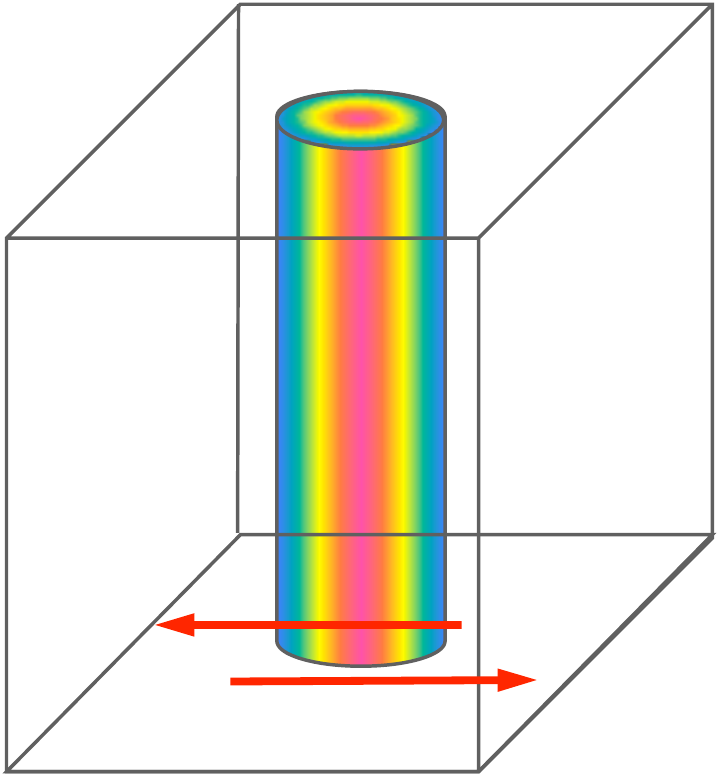}}
\scalebox{.18}{\includegraphics{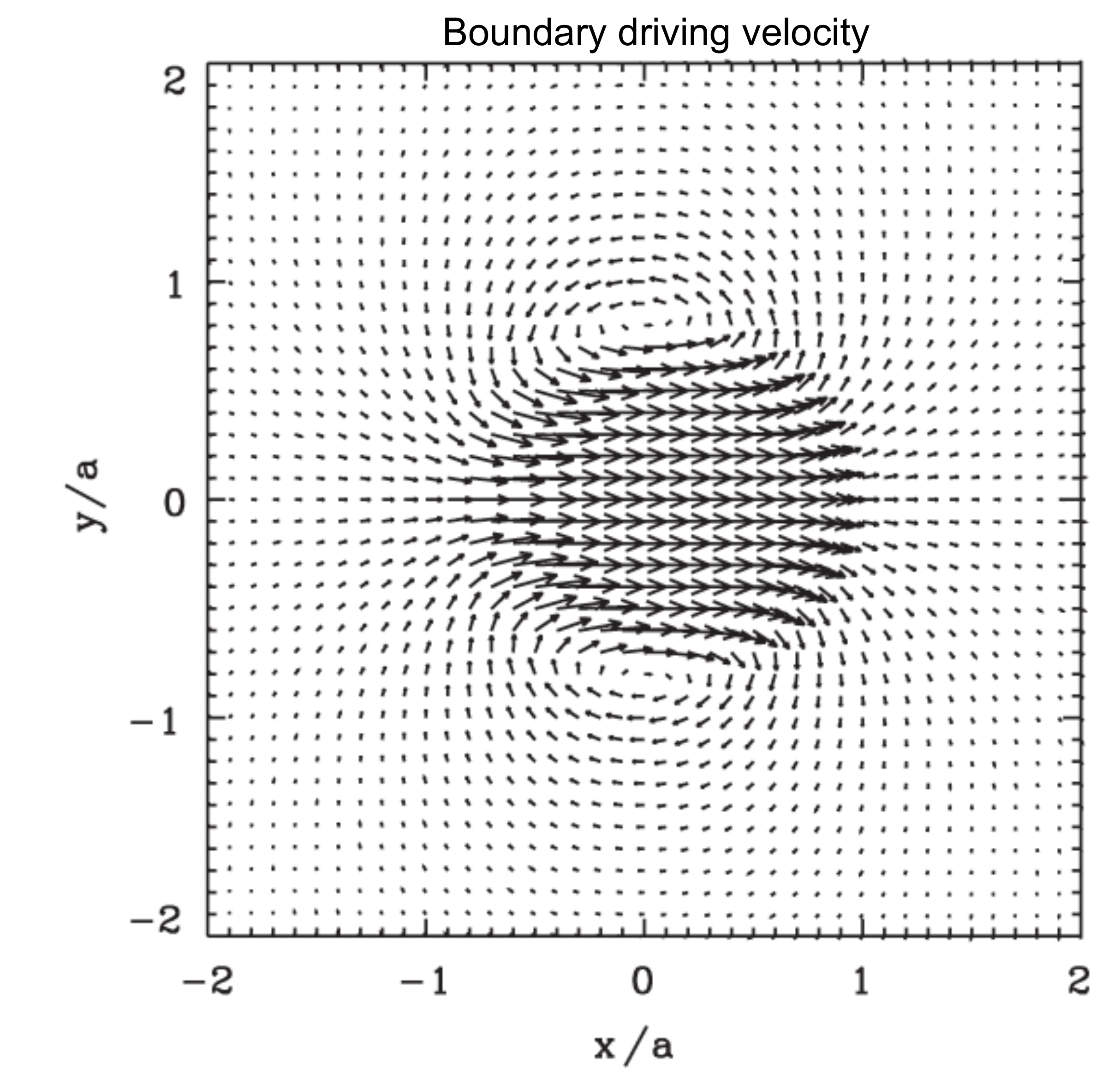}}}
\caption{This figure describes the basic setup of the 3D numerical simulations. (left) A cross-section through the horizontal (transverse) density structure of the flux tube. (middle) Cartoon showing the flux tube density structuring (red=higher density) and the basic footpoint displacement in the 3D numerical domain. (right) A snapshot of the 2D-dipole velocity driver on the lower $z$-boundary.}
\label{fig:setup}
\end{figure}

The basic setup of the 3D numerical model is described in detail in \cite{Pascoe2010} and summarised in Figure \ref{fig:setup}: a cylindrical density enhancement is embedded in a vertical, uniform magnetic field (see middle panel of Figure \ref{fig:setup}). The density enhancement is defined as a cylindrical core region of radius $r \le b$ with constant density $\rho_0$, surrounded by an inhomogeneous layer $b < r \le a$ where the density decreases linearly from $\rho_0$ to $\rho_e$, the external density. For the example shown in Figure \ref{fig:setup}, the density contrast is $ \rho_{0} / \rho_{e} = 2 $ and  the inhomogeneous layer has thickness $l/a=0.5$, where $l=a-b$. As the core region defines a minimum in the Alfv\'en speed, it forms a wave-guide for magneto acoustic waves (e.g.~\cite{Edwin1983}, \cite{Roberts1984}).

A velocity driver, consisting of only horizontal (i.e.~$x$ and $y$) components, is applied at the lower ($z$) boundary.  The driver is chosen to simulate generic solar surface footpoint motions and consists either of a single pulse (displacement) or continuous harmonic footpoint motions. The basic motion is a 2D (horizontal) dipole motion as described in \cite{Pascoe2010} and shown in the RHS panel of Figure \ref{fig:setup}, with an amplitude chosen sufficiently small (of the order of 0.2\% of the external Alfv\'en speed) to avoid non-linear effects and describe the regime relevant to the CoMP observations. In the core of the flux tube (constant-density region), the velocity is constant and in the $x$-direction only. In cylindrical coordinates, this 2D incompressible dipole flow would correspond to the $m=1$ mode.

The simulations are performed using the 3D MHD code \textsc{Lare3D} (\cite{Arber2001}). The boundary conditions are periodic in the $x$ and $y$ directions, and are placed sufficiently far from the flux tube to not affect the results. The resolution is chosen to be much higher in the $x$ and $y$ directions in order to resolve the small-scale wave motions (due to phase mixing) in the flux tube shell region for as long as possible.

\section{Mode coupling to the azimuthal Alfv\'en wave}

\begin{figure}[t]
\centering{
\scalebox{.4}{\includegraphics{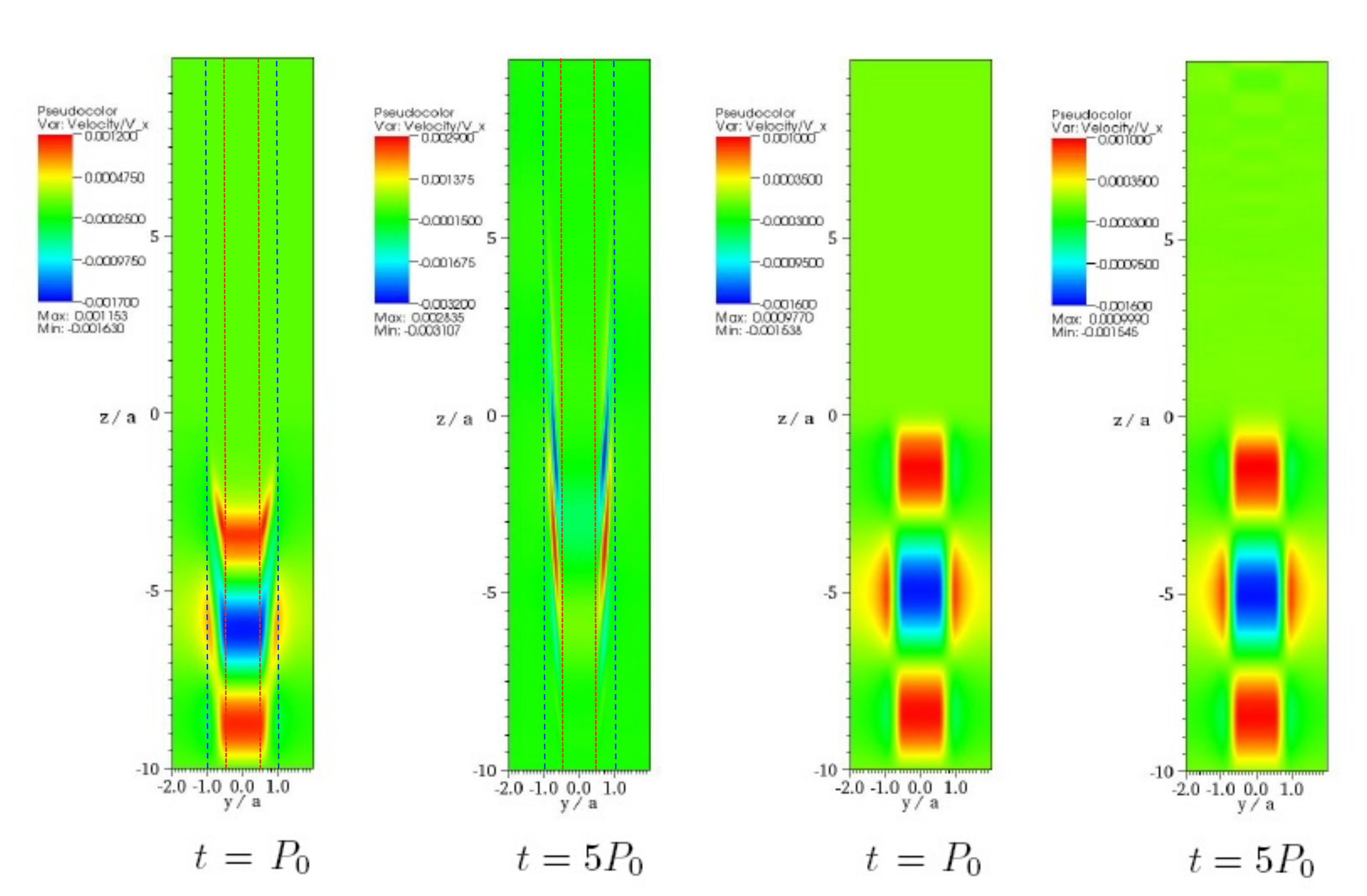}}}
\caption{(left panels) A vertical cross-section of the transverse velocity component $v_x$ (at an early time $t=P_0$ and a later time $t=5P_0$). The red vertical lines outline the core of the fluxtube (where the density is constant) and the shell region is between the red and blue lines. (right panels) The same results but from a simulation where the density in the 3D numerical domain is constant (i.e.~no fluxtube is present).}
\label{fig:Alfvenic}
\end{figure}

The results of the basic numerical simulations are shown in the two panels on the LHS of Figure \ref{fig:Alfvenic}. The first panel shows a vertical cross-section (i.e.~along the $z$-axis) through the numerical domain of the transverse velocity component $v_x$, at a time $t=P_0$ (where $P_0$ is the period of the boundary driven pulse) and the second panel corresponds to a later time $t= 5 P_0$. We note here that for computational efficiency, the numerical domain was made periodic in the vertical direction once the boundary driven pulse had propagated fully into the domain. The vertical red lines outline the core of the flux tube where the density is constant and the shell (or boundary) layer is situated between the red and blue lines. 

Comparing the LHS panels in Figure \ref{fig:Alfvenic}, the mode coupling process is clearly evident: where the phase speed of the (driven) kink wave packet matches the local Alfv\'en speed (in the shell region), energy is transferred very efficiently to the azimuthal ($m=1$) Alfv\'en wave in the boundary layer. In other words, the driven kink wave propagating along the loop acts as a moving source of Alfv\'en waves and the efficient energy transfer into the boundary layer implies that the amplitude of the (observable) transverse velocity perturbations (the kink wave) in the core of the loop decays within a few wave periods, consistent with the observational results of \cite{Tomczyk2009}. The efficiency of the energy transfer, or in other words, the rate at which the central, transverse velocity perturbations is seen to decay, depends on the boundary layer thickness, the density contrast between the core of the loop and the surrounding plasma and the frequency of the incoming (boundary-driven) perturbation. Damping is more efficient for wider boundary layers, higher density contrast and higher frequencies (see \cite{Pascoe2010} for a comprehensive description of these results).

As a comparison, the RHS panels in Figure \ref{fig:Alfvenic} show exactly the same velocity component, at the same times, but for a simulation where the density was assumed uniform (corresponding to the external density) everywhere. The same 2D dipole driver was applied to the lower $z$-boundary and it is obvious that in this uniform medium no mode coupling takes place and the transverse velocity perturbation (now a shear Alfv\'en wave) simply propagates unmodified.

At this point, it is worthwhile reflecting on the nature of the ``$m=1$ Alfv\'en mode'' in more detail. The $m=1$ Alfv\'en wave results from the resonant coupling of the fast and Alfv\'en wave solutions. When $m=0$, these modes decouple exactly, but when $m \ne 0$, the situation becomes more complex. Strictly speaking, when $m=1$, there is no fast or Alfv\'en wave solution, but  a coupled set of fields that satisfies the mathematical equations. However, in certain regions, these fields are predominantly fast or Alfv\'enic in character. It is in this spirit that we use the term ``$m=1$ Alfv\'en wave''. There is good justification for this: the field-aligned structure and period of the $m=0$ and $m=1$ Alfv\'en waves are identical and both are (essentially) incompressible. Indeed, even though an $m=1$ $B_{\phi}$ field by itself cannot satisfy $\nabla \cdot {\bf B} =0$ (and a $v_{\phi}$ velocity component alone will compress the plasma to some extent), \cite{Wright1992} showed that when the fields have a small scale in the radial direction (either through being a resonant solution or through being strongly phasemixed), the solution is essentially incompressible: only negligibly small $B_{r}$ ($\ll B_{\phi}$) and even smaller $B_{z}$ components are needed to satisfy $\nabla \cdot {\bf B} =0$, which to leading order balances $B_r/\delta r \approx m B_{\phi}/r$ ($\delta r$ being the small radial scale of the wave). A clear discussion in a Cartesian formulation is given by \cite{Mann1997} for the large $m$ limit, where an analogous problem occurs. Hence, the term ``$m=1$ Alfv\'en wave'' is appropriate for the leading order perturbations which are decoupled and incompressible Alfv\'enic fields in an asymptotic (rather than exact) sense.

\section{Gaussian Damping}

\begin{figure}[t]
\centering{
\scalebox{.2}{\includegraphics{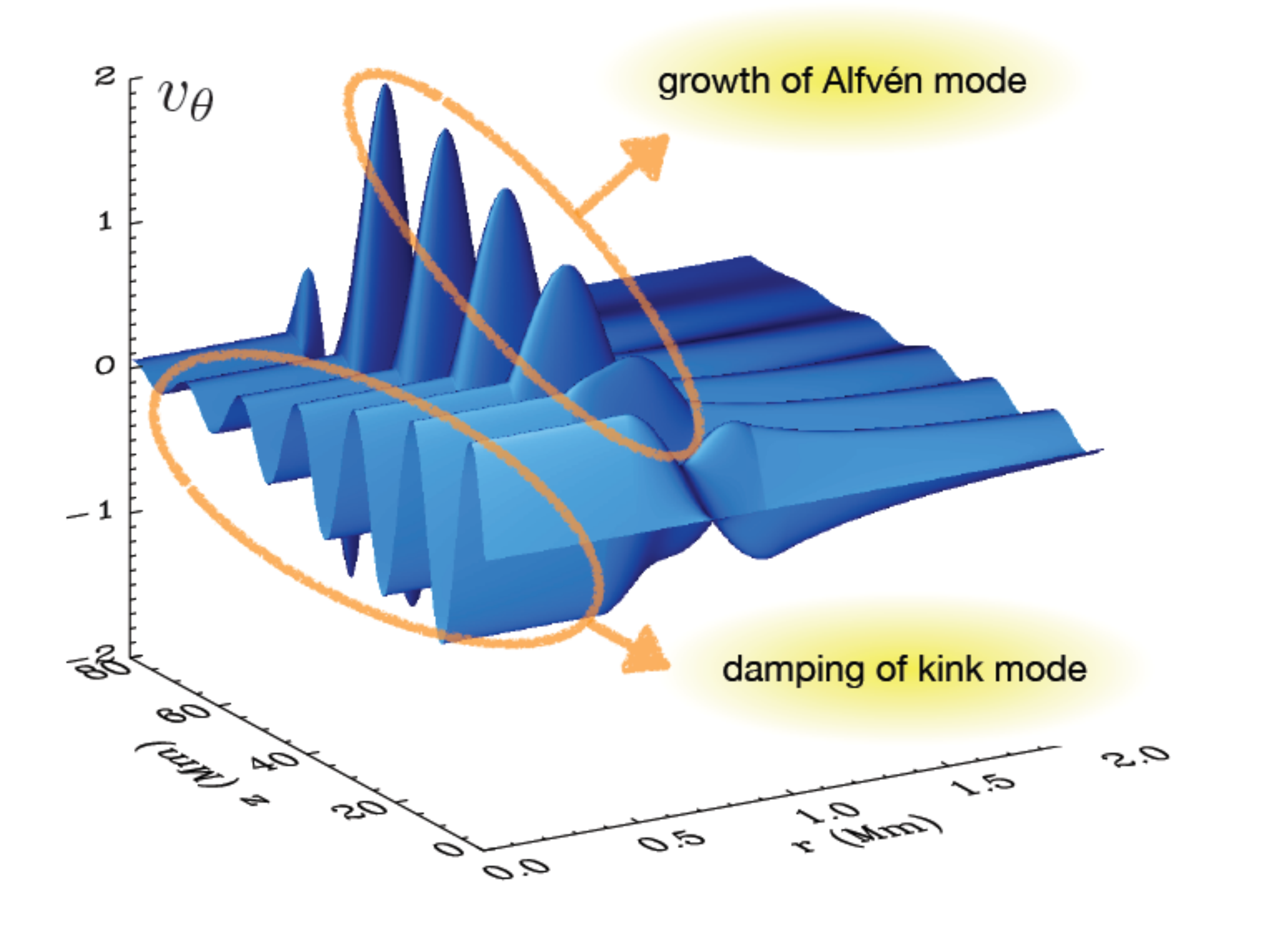}}
\scalebox{.2}{\includegraphics{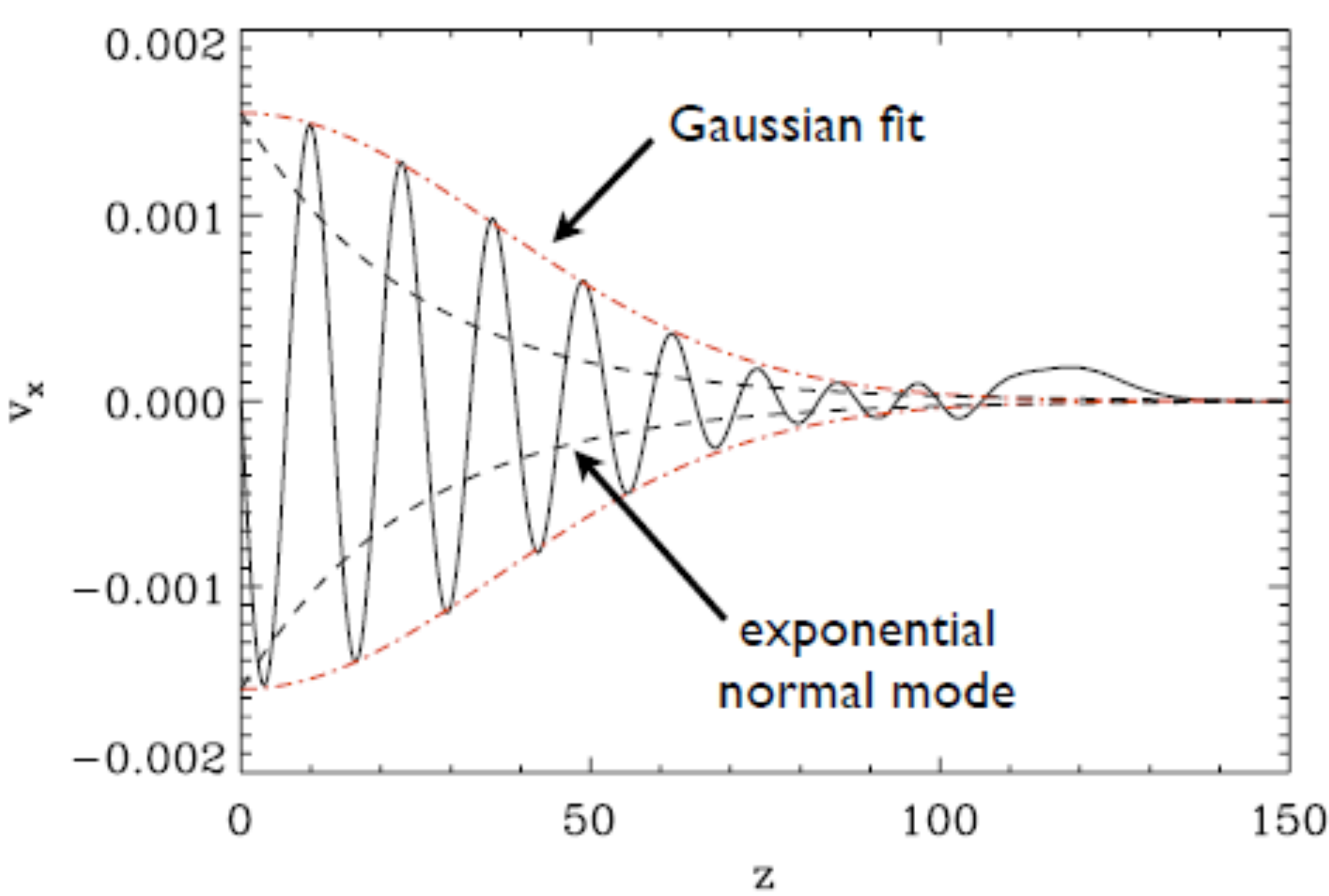}}
\scalebox{.169}{\includegraphics{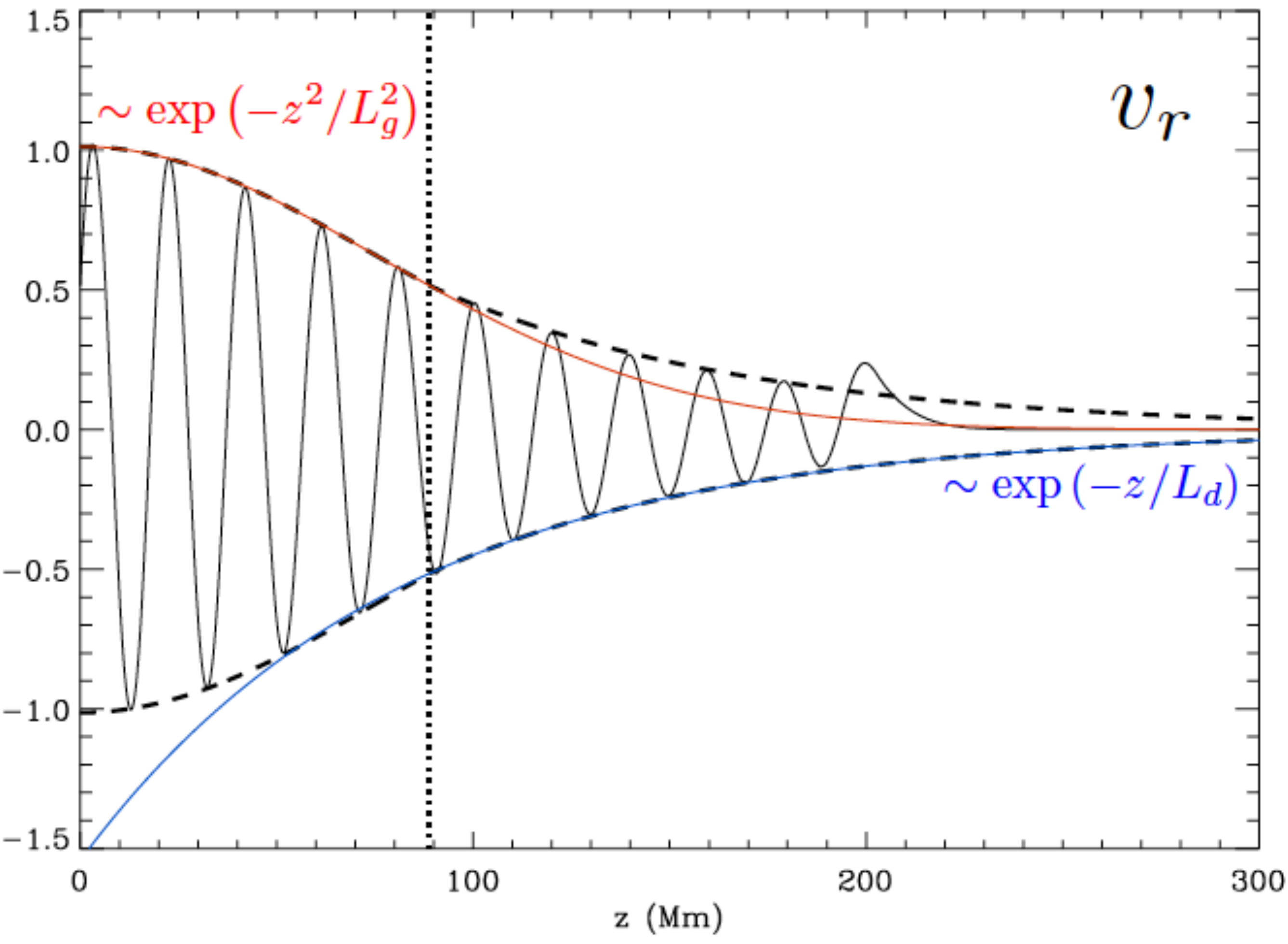}}}
\caption{(left) A surface plot of the azimuthal velocity $v_\theta$ along the line $\theta=0$. (middle/right) A cross-section of the transverse velocity (middle) $v_x$ or (right) $v_r$ at the centre of the fluxtube as a function of height.}
\label{fig:Gaussian}
\end{figure}

As energy is transferred to the boundary Alfv\'en wave due to the mode coupling, the amplitude of the bulk, transverse velocity perturbation in the centre of the flux tube decays. The LHS panel of Figure \ref{fig:Gaussian} shows the azimuthal component of the perturbed velocity ($v_\theta)$ along the $\theta=0$ line and the growth of the azimuthal Alfv\'en wave in the boundary layer is clearly visible. Note that these simulations were performed in cylindrical coordinates. The central panel of Figure \ref{fig:Gaussian}  shows a cross-section of the radial velocity component $v_r$ at the centre of the fluxtube ($r=0$), which is a signature of the kink mode. It is clear that the transverse velocity damps rapidly as a function of the height $z$. Overplotted in the black dashed lines is the exponential damping profile, $\exp(-z/L_d)$ (where $L_d$ is the damping length), suggested by (traditional) normal mode analysis. It is obvious that this damping profile does not match the numerical simulations and \cite{Pascoe2012} demonstrated that the damping instead follows an (empirical) Gaussian profile (overplotted in red dashed lines) of the form $\exp(-z^2/L^2_g)$, where the damping length $L_g$ does not necessarily match $L_d$. 

Following the empirical determination of the Gaussian damping by \cite{Pascoe2012}, the unexpected damping profile was confirmed with detailed analytical calculations by \cite{Hood2013} and an extensive (numerical) parameter study by \cite{Pascoe2013}. The analytical study suggested that the Gaussian damping profile was most appropriate at lower heights, which allowed \cite{Pascoe2013} to construct a general damping profile for the amplitude $A(z)$ of the perturbed, transverse velocity, of the form
\begin{equation}
A(z)=\left\{\begin{array}{cc}
\frac{A_0}{2} \left[1+\exp\left( -\frac{z^2}{L^2_g}\right) \right]  \qquad & z\le h,\\
A_h \exp\left( - \frac{z-h}{L_d}\right) & z>h,
\end{array}
\right.
\end{equation}
where $A_h=A(z=h)$ and the height of the switch between the two profiles is given by $h=L^2_g/L_d$. This height $h$, where the switch between the two profiles occurs, was found to depend only on the density contrast between the core of the loop and the surrounding plasma. The two damping lengths are given by
\begin{equation}
L_g =  \frac{16}{\epsilon k^2 \kappa^2}\qquad \textrm{and} \qquad L_d=\frac{8}{\pi \epsilon k \kappa}
\end{equation}
where $k$ represents the wavenumber, $\epsilon$ corresponds to the thickness of the loop shell region (normalised with respect to the loop radius) and the parameter $\kappa=(\rho_0 - \rho_e)/ (\rho_0 + \rho_e)$ is a measure of the contrast between the density $\rho_0$ of the loop core and the density $\rho_e$ of the external plasma. It is clear that any seismological inversions using the observed damping length would lead to a different outcome, depending on whether the Gaussian or exponential damping profile is applied to the observations. \cite{Pascoe2013} demonstrated that applying the Gaussian profile would be most appropriate for loops with a low density contrast or when only a few wavelengths (along the loop) are observed. 

\section{Discussion \& Conclusions}

The 3D numerical simulations demonstrated that when a dense flux tube is present (with a smooth transverse/radial density profile), no matter how weak the actual density contrast with the surrounding plasma, the boundary driven transverse perturbations (``kink'' modes) will couple to azimuthal Alfv\'en waves in the loop shell region. Both of these waves are dominated by velocity perturbations that are transverse to the magnetic field. Hence, the interpretation of observed Doppler velocity perturbations propagating along coronal loops in terms of either kink or Alfv\'en waves might just not be possible. The more general term ``transverse wave perturbations'' appears most appropriate (even more so when more generic 3D models are considered, see e.g.~\cite{Ofman2009} for examples of 3D simulations).

The numerical simulations of \cite{Pascoe2010}, \cite{Pascoe2012} and \cite{Pascoe2013} used a cylindrically symmetrical density enhancement and a 2D dipole-like boundary driver with a radius close to the loop radius. However, \cite{Pascoe2011} demonstrated that although transverse density structuring clearly has to be present to allow the mode coupling to take place, this structuring does not have to be regular (i.e.~cylindrically symmetric) and the footpoint motion does not necessarily have to coincide exactly with the density enhancement which forms the loops. Mode coupling will occur at any location where the phase speed of the driven kink waves matches the local Alfv\'en speed. In other words, the process of mode coupling is a very robust (or unavoidable) one and likely to take place in many coronal structures. Again breaking the symmetrical and co-aligned nature of the driver and the flux tube, \cite{Pascoe2015} implemented a boundary driver mimicking small-scale turbulent motions. These eddies, with spatial scales smaller than the fluxtube width, still excite propagating kink waves (which will mode-couple to azimuthal Alfv\'en waves) but were much less efficient than the larger-scale, 2D-dipole driver. Hence, these authors suggested it was more likely that the observed Alfv\'enic waves in coronal loops were driven by footpoint motions on spatial scale at least comparable to the loop width.

Finally, \cite{Pascoe2015} considered a broadband-frequency footpoint driver. Previous studies (e.g.~\cite{Ruderman2002}, \cite{Pascoe2011}, \cite{Terradas2010}, \cite{Verth2010}) showed the damping length of the kink waves to be proportional to the wave period ($L_d \sim P)$, or, in other words, that mode-coupling will have a frequency filtering effect, where the high frequency modes are expected to damp faster and the longer period waves are expected to propagate further. \cite{Pascoe2015} describe a method which does not make a priori assumptions for the damping behaviour but allows seismological information to be inferred from measurements of the frequency-dependent damping rate of broadband kink waves. Unfortunately, for the perturbations observed by \cite{Tomczyk2007}, the signal-to-noise was too low to be able to distinguish between the different forms of frequency-filtering associated with the two different spatial damping profiles (\cite{Pascoe2015}). In addition, further observations by CoMP (\cite{IDM2014}, \cite{Liu2014}) found an excess of high-frequency wave power in long coronal loops compared to the expected damping rates, suggesting additional effects such as the enhancement of high frequencies by turbulent cascade might be present. Recent modelling (e.g.~\cite{VanBallegooijen2011}) showed that such wave turbulence could be caused by the non-linear interactions of counter-propagating waves. (Note however that \cite{VanBallegooijen2011} use incompressible, reduced MHD and hence cannot model the coupled kink and Alfv\'en waves.) For the coronal loops observed by CoMP, counter-propagating waves could consist of two wave trains traveling upwards from two opposing footpoints or the fast, transverse (Doppler shift) disturbances could be interacting with slower propagating density perturbations, possibly leading to reflections and hence, counter-propagating waves. The co-existence of such slow propagating density perturbations and the faster transverse waves was recently observed by e.g.~\cite{Threlfall2013} in coronal loops and \cite{Liu2015} in coronal plumes.

In summary, recent observations have demonstrated that transverse waves and oscillations are ubiquitous in the solar atmosphere and potentially contain a substantial amount of energy.  Mode coupling between propagating, transverse velocity perturbations (MHD kink waves) and azimuthal Alfv\'en waves is a very robust process which can qualitatively account for the observed rapid damping of many of the observed transverse perturbations (Doppler shifts). Numerical simulations have shown that the spatial damping profile of such waves will be Gaussian in nature for at least the first few wavelengths. We note here that the Gaussian damping behaviour also applies to the time variation of standing kink modes at early times  (\cite{Pascoe2013}, \cite{Ruderman2013}). Finally, we point out that although observations have now shown unambiguously that waves are present in the solar atmosphere, this does not automatically imply a solution to the coronal heating process or the acceleration of the solar wind; the process of mode coupling simply transfers the energy into a different wave mode. It remains to be demonstrated (theoretically or with numerical modelling) whether the subsequent dissipation of the azimuthal Alfv\'en waves, enhanced by e.g.~phase mixing or the Kelvin-Helmholtz instability in the inhomogeneous fluxtube boundaries, could lead to local plasma heating on the appropriate timescales (see e.g.~\cite{Hollweg1988}, \cite{Ofman1994}, \cite{Terradas2008} or \cite{Antolin2014}).
 
\ack
The computational work for this paper was carried out on the joint STFC and SFC (SRIF) funded cluster at the University of St Andrews (Scotland, UK). The research leading to these results has also received funding from the European Commissionâ Seventh Framework Programme (FP7/2007-2013) under the grant agreement SOLSPANET (project No. 269299,  www.solspanet.eu/solspanet).

\section*{References}
\bibliographystyle{jphysicsB}
\bibliography{Rev_ModeCoupling}

\end{document}